\documentclass[aps,prb,twocolumn,superscriptaddress,showpacs]{revtex4}
\usepackage{graphics}
\usepackage{epsfig}
\usepackage{graphicx}
\usepackage{dcolumn}
\usepackage{bm}
\usepackage{graphicx}
\usepackage{dcolumn}
\usepackage{rotating}
\usepackage{supertabular}
\usepackage{amssymb}
\usepackage{amsmath,theorem}
\usepackage{threeparttable}
\usepackage{multirow}

\newlength{\myleno}
\newlength{\mylent}
\newcommand{\fr}{\ensuremath{\mathbf{r}}}

\begin{document}

\title{Accurate Hartree-Fock energy of extended systems using large Gaussian 
basis sets}

\author{Joachim Paier}
\affiliation{Department of Chemistry, Rice University, Houston, Texas 7705, USA}

\author{Cristian V. Diaconu}
\affiliation{Department of Chemistry, Rice University, Houston, Texas 7705, USA}

\author{Gustavo E. Scuseria}
\affiliation{Department of Chemistry, Rice University, Houston, Texas 7705, USA}

\author{Manuel Guidon}
\affiliation{Physical Chemistry Institute, University of Zurich, Winterthurerstrasse 190, 
CH-8057 Zurich, Switzerland}

\author{Joost VandeVondele}
\affiliation{Physical Chemistry Institute, University of Zurich, Winterthurerstrasse 190, 
CH-8057 Zurich, Switzerland}

\author{J{\"u}rg Hutter}
\affiliation{Physical Chemistry Institute, University of Zurich, Winterthurerstrasse 190, 
CH-8057 Zurich, Switzerland}

\date{\today}

\begin{abstract}
Calculating highly accurate thermochemical properties of
condensed matter via wave function-based approaches (such as {\it e.g.}~Hartree-Fock or
hybrid functionals) has recently attracted much interest. 
We here present two strategies providing accurate
Hartree-Fock energies for solid LiH  
in a large Gaussian basis set and applying periodic boundary conditions. 
The total energies were obtained using two
different approaches, namely a supercell evaluation of
Hartree-Fock exchange using a truncated Coulomb operator 
and an extrapolation toward the full-range Hartree-Fock limit of 
a Pad{\'e} fit to a series of short-range screened Hartree-Fock calculations.
These two techniques agreed to significant precision.
We also present the Hartree-Fock cohesive energy of LiH (converged to within sub-meV) 
at the experimental equilibrium
volume as well as the Hartree-Fock equilibrium lattice constant and bulk modulus.
\end{abstract}

\pacs{61.50.Ah, 71.15.-m, 71.15.Nc, 71.15.Ap}

\maketitle
%
\section{Introduction\label{sec:intro}}
The high accuracy/cost ratio of Kohn-Sham density functional theory\cite{kohn_sham:pr:65,dreizler} (KS-DFT)
has been exhaustively demonstrated in the literature. In its early days, KS-DFT
using the local density approximation\cite{kohn_sham:pr:65} was almost exclusively applied by the solid
state community. However, the advent of generalized gradient approximations 
(GGAs, see {\it e.g.}~Refs.~\onlinecite{langreth:prb:80,perdew:prb:92,perdew:prl:96})
to the exchange-correlation (XC) functional and the introduction of nonlocal Hartree-Fock (HF) exchange in
 hybrid functionals\cite{becke:jcp:92,becke:jcp:93} paved the way for reasonably accurate applications
to molecules as well.

Within the framework of KS-DFT it is relatively easy to achieve basis set convergence, and atomic forces
can be calculated at little extra computational cost. This is of paramount importance in the calculation of
high-temperature dynamical and thermodynamic properties by molecular dynamics simulations.
In particular, DFT statistical mechanics for both bulk materials and for surface processes is
routinely feasible (see {\it e.g.}~Ref.~\onlinecite{aimd}, and
references therein). The principle limitation of KS-DFT lies in the accuracy of the applied XC functional.

Discussing examples for some shortcomings of KS-DFT, 
it is well known that standard local and semilocal approximations to the XC functional do not yield accurate results
for quasiparticle band-gaps of semiconductors and insulators. Generally, they do not predict the correct adsorption
sites and adsorption energies of molecules on metallic surfaces 
(for details see {\it e.g.}~Ref.~\onlinecite{stroppa:njp:08}, and references therein).
Today's DFT practitioner is confronted with these shortcomings when choosing an XC functional for a specific
application. Each contemporary density functional has its relative merits but at the same time
drawbacks, which might impede finding an appropriate XC functional. 
Different density functionals have different merits and demerits, an unsatisfactory situation. 
These inadequacies of semilocal KS-DFT have stimulated some DFT groups to use 
wave function-based methods to benchmark or correct DFT results.
Note that another large community of researchers directly apply wave function-based techniques
to materials science problems (see Ref.~\onlinecite{paulus:pr:06}, and references therein).

Dramatic improvements for many properties of molecules as well as solids can be
achieved by mixing a fraction of nonlocal HF exchange (HFX) to the remaining part
of semilocal DFT exchange. Since these functionals do not only depend on
the electron density alone, but also on the KS single particle wave functions, {\it i.e.}~the orbitals,
 they are called \emph{hybrid} functionals. Therefore, these hybrid functionals can be seen as
``mixed'' wave function-based and semilocal DFT methods.
We refer the reader to a recent review~\cite{janesko:pccp:09} of so-called screened hybrid functionals, 
as {\it e.g.}~the Heyd-Scuseria-Ernzerhof\cite{heyd:jcp:03,heyd:jcp:06}
(HSE) functional, which was proposed to extend the successes of
hybrid functionals into condensed matter, by avoiding the problematic
effects of long-range HFX (see Ref.~\onlinecite{janesko:pccp:09}, and
references therein).

Besides the successes of screened HFX applied to condensed matter, 
the numerous methodological and algorithmic developments in the quantum-chemistry community and the steady
increase of computers' efficiency induced a drive to conceive and implement even more involved
wave function-based techniques, as {\it e.g.}~local second-order M{\o}ller-Plesset perturbation theory 
(MP2)~\cite{pisani:jcp:05,pisani:jcc:08,casassa:tca:07}, (resolution of
the identity) atomic orbital Laplace transformed MP2~\cite{ayala:jcc:00,ayala:jcp:01,izmaylov:pccp:08} and canonical
MP2~\cite{sun:jcp:96,marsman:jcp:09} for (infinitely) extended systems of various dimensionality and applied basis functions.
Furthermore, recent reports in the literature on {\it ab initio} molecular dynamics on 
condensed matter\cite{guidon:jcp:08} employing the HSE screened hybrid functional illustrate that, 
depending on implementation details, basis set and system, wave function-based techniques are also applicable
to statistical mechanics calculations. 
These successful applications of wave function methods to large systems
show that they are able to tackle materials science problems
with possibly much better accuracy than conventional density functionals.

Recently published HF and post-HF calculations on crystalline LiH
have attracted much interest in the solid state community\cite{manby:pccp:06,gillan:jcc:08,nolan:prb:09}.
 These calculations represent a benchmark in terms of eliminating as many inaccuracies as possible while
attempting to converge toward the so-called HF limit.
The approach in question employs calculations on a hierarchical series of cluster models,\cite{manby:pccp:06,nolan:jphyscond:09} 
exploiting strengths and weaknesses of plane wave pseudopotentials as well as local Gaussian basis sets.
Accurate evaluation of the total HF energy, as well as cohesive energy in the HF approximation employing exclusively
Gaussian basis sets is desirable to bypass errors incurred by the pseudopotential approximation.
Admittedly, creating an all-electron Gaussian basis set, which describes the crystal as well as the
isolated atoms equally well, is challenging.
Referring to the arguments of Gillan {\it et al.},~\cite{gillan:jcc:08} it is in general difficult
to provide rigorous estimates how far the applied basis set is from the HF limit. However, it
is reasonable to question the need for reaching the HF limit for particular materials properties, which
is substantiated in the present work.

We compare total HF energies of solid LiH using two different codes employing
Gaussian basis functions: 
(i) the Gaussian and augmented-plane wave (GAPW)~\cite{parrinello:tca:99} 
code CP2K/Quickstep~\cite{vandevondele:cpc:05,cp2k}
and (ii) a developmental version of the
GAUSSIAN suite of programs~\cite{gaussian}.
We show that the cohesive energy of the crystal is converged to within sub-meV
accuracy in our given large Gaussian basis set (see Tab.~\ref{tab:basis}).
Computational and methodological details are presented in Sec.~\ref{sec:comp}.
Results for cohesive energies, theoretical lattice constant as well as bulk modulus are in
Sec.~\ref{sec:res}.
Conclusions are drawn in Sec.~\ref{sec:conc}.

\section{Computational details\label{sec:comp}}
In the following sections, we describe important computational details,
 such as the Gaussian basis set, the evaluation of full-range HFX based on the 
short-range (SR) HFX implementation~\cite{izmaylov:jcp:06} in the GAUSSIAN suite of programs
as well as the method applied for the extrapolation of the SR-HFX energy to the full range limit based on 
Pad{\'e} approximants. Furthermore, implementation details on the direct evaluation of HFX
via the CP2K/Quickstep code are presented.

\subsection{Basis set}
The basis set used for this calculation has been specifically constructed for the current purpose, which is
an accurate but computationally feasible HF calculation on bulk LiH.
The basis constructed here is similar to the polarization consistent (pc) basis sets derived 
by Jensen.\cite{jensen:jcp:01,jensen:jcp:02,jensen:jpc:07}
Jensen introduced a sequence of quasi-optimal basis sets (pc-[0-4]) that rapidly converge to the HF and DFT basis set limit.
The pc-3 basis set gives atomization energies with a mean error smaller than 1 kJ/mol.
For H and Li the pc-3 basis set has a composition 9s4p2d1f/5s4p2d1f and 14s6p2d1f/6s3p2d1f respectively,
while we adopt 8s3p2d1f/6s3p2d1f and 13s6p2d1f/11s5p2d1f.
However, the primitives of the basis employed here are non-standard and optimized for the present calculations.

In a first step, we have removed primitive Gaussians with exponents smaller than $0.15~\mathrm{bohr}^{-2}$, since
diffuse basis functions are technically troublesome. Diffuse functions, which are needed to describe density tails
in atoms or molecules, are not needed in the bulk of densely-packed solids with large band gaps as the case of LiH. 
Indeed, we exploit the fact that the basis functions on the lattice sites are available for the expansion of any orbitals,
be it the crystal orbitals in the bulk or the atomic orbitals of the isolated atoms.
This basis is thus only suited for atomic or surface calculations if ghost basis functions are left 
in the regular lattice positions to appropriately describe the 
aforementioned tails of the electron density.

In a second step, all but the core exponents have been optimized by minimizing the energy of bulk LiH subject
to a restraint on the condition number of the overlap matrix.
This procedure is similar to the one employed for the molecularly optimized basis sets described in 
Ref.~\onlinecite{vandevondele:jcp:07}.
In CP2K, density functionals that do not include Hartree-Fock exchange can be computed in a highly efficient manner,
and in order to make this procedure computationally efficient, such a semilocal density functional 
(B88~\cite{becke:pra:88}) has been employed in the
optimization process.
The resulting basis is well conditioned, the condition number of the overlap matrix is $2.8\times 10^4$ for bulk LiH.
We have estimated the accuracy of the optimized basis by comparing to pc-4-like basis sets,
which for this system are only feasible with local DFT,
and estimate the total energy of bulk LiH (per unit of LiH) to be well within 0.001 a.u. of the basis set limit,
while the basis set error on the cohesive energy is likely smaller than 0.1\% (0.0001 a.u.).
The details of this optimized basis set are summarized in Tab.~\ref{tab:basis}.

\begin{table}[tb]
\renewcommand{\arraystretch}{1.1}
\caption{\label{tab:basis} Details for the adopted basis sets for the compositions 8s3p2d1f/6s3p2d1f and 
13s6p2d1f/11s5p2d1f of Hydrogen and Lithium respectively. Shown are angular momentum, Gaussian exponent 
and corresponding contraction coefficients.}
\begin{ruledtabular}
\begin{tabular}{llcc}
species  &  l     & exponent   &  coefficient   \\
\hline
H  &  s   & 0.27463675e{02}  & 1.00000000   \\
   &  s   & 0.68559258e{01}  & 1.00000000   \\
   &  s   & 0.17679972e{01}  & 1.00000000   \\
   &  s   & 0.51181842e{00}  & 1.00000000   \\
   &  s   & 0.20167548e{00}  & 1.00000000   \\
   &  s   & 0.30797000e{04}  & 0.00023473   \\
   &      & 0.46152000e{03}  & 0.00182450   \\
   &      & 0.10506000e{03}  & 0.00959330   \\
   &  p   & 0.21240865e{01}  & 1.00000000   \\
   &  p   & 0.10736812e{01}  & 1.00000000   \\
   &  p   & 0.56838662e{00}  & 1.00000000   \\
   &  d   & 0.92833840e{00}  & 1.00000000   \\ 
   &  d   & 0.49583000e{00}  & 1.00000000   \\
   &  f   & 0.12073480e{01}  & 1.00000000   \\
Li &  s   & 0.13360341e{04}  & 1.00000000   \\
   &  s   & 0.44429982e{03}  & 1.00000000   \\
   &  s   & 0.14779702e{03}  & 1.00000000   \\
   &  s   & 0.49209451e{02}  & 1.00000000   \\
   &  s   & 0.16428957e{02}  & 1.00000000   \\
   &  s   & 0.55293994e{01}  & 1.00000000   \\
   &  s   & 0.19052824e{01}  & 1.00000000   \\
   &  s   & 0.70025874e{00}  & 1.00000000   \\
   &  s   & 0.29958682e{00}  & 1.00000000   \\
   &  s   & 0.16636288e{00}  & 1.00000000   \\
   &  s   & 0.70681000e{05}  & 0.00000544   \\
   &      & 0.13594000e{05}  & 0.00003328   \\
   &      & 0.31004000e{04}  & 0.00019175   \\
   &  p   & 0.15709110e{01}  & 1.00000000   \\
   &  p   & 0.74875864e{00}  & 1.00000000   \\
   &  p   & 0.38614089e{00}  & 1.00000000   \\
   &  p   & 0.22620503e{00}  & 1.00000000   \\
   &  p   & 0.28500000e{02}  & 0.00036754   \\
   &      & 0.66400000e{01}  & 0.00322359   \\
   &  d   & 0.77920820e{00}  & 1.00000000   \\
   &  d   & 0.40789925e{00}  & 1.00000000   \\
   &  f   & 0.73706300e{00}  & 1.00000000   \\
\end{tabular}
\end{ruledtabular}
\end{table}

\subsection{Extrapolation of SR-HFX to full range (GAUSSIAN)\label{sec:srHF}}

All Gaussian calculations presented in this work are based on a very efficient implementation
of the  SR-HFX energy exploiting a distance based screening protocol.\cite{izmaylov:jcp:06} 
Using local basis functions it is convenient to express the HFX energy for closed-shell as  
\begin{equation}
\label{eq:def-hf}
E_x^{\rm HF} = -\frac12 \sum_{\mu\nu\lambda\sigma} P_{\mu\lambda}P_{\nu\sigma} (\mu\nu|\lambda\sigma)_g,
\end{equation}
where $P_{\mu\nu}$ are density matrix elements and
\begin{equation}
(\mu\nu|\lambda\sigma)_g = \int  \mu(\fr_1)\nu(\fr_1) g(r_{12}) 
\lambda(\fr_2)\sigma(\fr_2)\,d\fr_1 d\fr_2\,
\end{equation}
are the four-center electron repulsion integrals (ERIs), represented in an atomic orbital basis. 
The applied interaction potential $g(r_{12})$ is usually equal to the Coulomb kernel $\frac{1}{r_{12}}$.

For large gap systems, it has been shown that local single particle wave functions
as well as the corresponding density matrix decay like $e^{-h |\fr_1 - \fr_2|}$ for large 
$|\fr_1 - \fr_2|$, where
$h$ is proportional to $\sqrt{\rm E_{\rm gap}}$, the square-root of the band gap of the system 
of question.\cite{kohn:pr:59,cloizeaux:pr:64,kohn:ijqc:95,goedecker:rmp:99}
This is the basic motivation behind SR-HF as {\it e.g.}~used 
in the successful HSE hybrid functional.\cite{heyd:jcp:03,heyd:jcp:06} 
HSE is based on a screened Coulomb interaction $g(r_{12})$ splitting the conventional Coulomb kernel, 
$\frac{1}{r_{12}}$, into 
\begin{equation}
\label{eq:scr-coul}
\frac{1}{r_{12}} = \underbrace{\frac{{\rm erfc}(\omega\, r_{12})}{r_{12}}}_{\rm SR} + 
\underbrace{\frac{{\rm erf}(\omega\, r_{12})}{r_{12}}}_{\rm LR},
\end{equation}
where the long-range (LR) and short-range (SR) parts of the interaction are described by the computationally  convenient
error function and its complement, respectively. The parameter $\omega$ in Eq.~\ref{eq:scr-coul} determines the extent
of the range separation of the Coulomb interaction.

In view of the relatively large HF band gap of LiH (10.8~eV)\cite{baroni:prb:85} and the fast decay of the density matrix,
we will calculate the total HF energy by doing a series of SR-HF calculations at different $\omega$, and extrapolating
to $\omega \rightarrow 0$.
As corroborated by numerical results shown in Sec.~\ref{sec:res}, such an extrapolation of the screened 
HF energies of the crystal to the full-range HF limit in the specified basis set is numerically robust and reliable.

All calculations are based on a locally modified development version of the GAUSSIAN electronic structure program.\cite{gaussian}
Hence, the total energies presented in Sec.~\ref{sec:res} do not include any DFT contributions. Only Hartree and
screened HFX energies are evaluated. The RMS convergence criterion for the density matrix in the self-consistent-field (SCF) iteration
 was set to $10^{-7}$~{\rm a.u.}, which implies an energy convergence no worse than at least
$10^{-8}$~{\rm a.u.} (GAUSSIAN keyword: SCF=Tight). 
Furthermore, a $24\times24\times24$ mesh of $k$ points was used, which is equivalent to 6912 $k$ points and thus all 
calculations are sufficiently converged with respect to $k$ points. 
The large band gap of LiH in the HF approximation (see above) substantially helps converging the $k$-point integration. 

Following ideas found in the literature,\cite{ayala:cpl:99,iyengar:ijqc:00}
we apply Pad{\'e} approximants of various orders to the obtained series of screened HF energies. The
actual form of the Pad{\'e} approximants are the rational polynomials
\begin{equation}
\label{eq:pade}
\frac{p(x)}{q(x)} = \frac{\sum_{i=0}^n p_i \, x^i}{\sum_{j=0}^m q_j\,x^j}.
\end{equation}
Eq.~\ref{eq:pade} represents the general expression of a Pad{\'e} approximant of order $[n/m]$.
Throughout this work only \emph{diagonal} rational polynomials are applied,\cite{ayala:cpl:99,num-rec}
which means that the order of the polynomial in the numerator equals the order of the polynomial
in the denominator.
Note that the number of parameters to be fitted is $2n +1 $ in the case of diagonal polynomials. 
This is the minimum number of data points, which must be included in the least-squares fit.

For all extrapolations employed in the present work, $\omega$ has been chosen to lie in the interval 
$[0.04\, ;\,1.0]$. In order to put a higher weight
to the area near to full-range HF we decided to increment $\omega$ by $0.005$ up to $0.1$ and increment $\omega$ by $0.01$
up to a value that amounts to $0.2$. 
For the remaining interval of larger $\omega$ values, the screening parameter was incremented by $0.1$.
As a consequence, each fit is based on 31 data points, representing pairs of the screening parameter $\omega$
and the corresponding SR-HF energy.

The HF equilibrium lattice constant and bulk modulus have been obtained by fitting the volume 
dependence of the static lattice energy to the Murnaghan equation of state.\cite{murnaghan:pnas:44} 
The points were chosen in order to cover a range of $\pm$3\% around the supposed equilibrium lattice constant
of 4.108~\AA\ (seven-points-fit).

\subsection{HFX and periodic boundary conditions using Gaussian basis functions\label{sec:hfx_gaussian}}
In hybrid functionals, which incorporate a
fraction of nonlocal HFX (Eq.~\ref{eq:def-hf}), the decay of the Hamiltonian matrix elements 
(see Sec.~II of Ref.~\onlinecite{kudin:prb:00}) with 
distance  is determined by two factors: 
(i) the decay behavior of the density matrix, $P_{\mu\nu}$, and (ii) the decay behavior of the ERIs.
For metallic as well as insulating systems, a screened Coulomb interaction 
accelerates the convergence of the ERIs in real-space drastically, {\it i.e.},
the number of replica cells needed for convergence is substantially decreased
(see Refs.~\onlinecite{heyd:jcp:04} and \onlinecite{paier:jcp:06}).
Fock exchange calculations involving the long-range tail of the Coulomb interaction 
({\it e.g.}~in the $\omega \rightarrow 0$ limit, see Eq.~\ref{eq:scr-coul}, in long-range corrected hybrids or
in global hybrids), both
the density matrix and the ERIs influence the convergence of the HFX energy.

It is a matter of fact, that due to the algebraic structure of Eq.~\ref{eq:def-hf}, 
contributions to the HFX energy can be significant 
even far from the central cell, precluding an early truncation of the lattice sum
(see Eq.~2.4 in Ref.~\onlinecite{kudin:prb:00}). 
Small exponent basis functions involved in the calculation of
the density matrix become important factors determining the computational workload.
Calculations under periodic boundary conditions (PBC) involving SR-HFX with a reasonably 
large value for the screening parameter $\omega$ (Eq.~\ref{eq:scr-coul}) are tractable for
moderately diffuse Gaussian basis functions, {\it i.e.}~minimal exponent equals $\approx\!0.2$.
Conventional HF or long-range HF calculations are likely to be computationally prohibitive
except for high-exponent Gaussian basis sets.
The relatively large and diffuse basis set used in this work (see Tab.~\ref{tab:basis})
prevents calculating the HFX at or close to $\omega = 0$, {\it i.e.}~the long-range limit for this particular
system in the given basis. As shown by the results presented in Sec.~\ref{sec:res}, a numerically stable
fit to a sufficiently large series of SR-HFX calculations is practicable to calculate an accurate
estimate for the HF energy of extended (insulating) systems using large Gaussian basis sets.
In summary, $\omega = 0$ is not practical whereas a 31 point $\omega$ extrapolation works very well.

\subsection{Direct calculation of HFX (CP2K)\label{sec:gpwHF}}
The focus of CP2K is the simulation of complex systems, with a variety of methods.
Recently, the capability to perform first principles molecular dynamics simulation 
with density functionals including a fraction of Hartree-Fock exchange has
been implemented and demonstrated for condensed phase systems containing a few hundred atoms~\cite{guidon:jcp:08}.
With this goal in mind, the implementation is massively parallel, focuses on in-core calculations, 
uses the $\Gamma$ point only,
and does not exploit molecular or crystal symmetries.
The implementation was based on a minimum image (MI) convention~\cite{tymczak:jcp:05} and employed a standard 1/r Coulomb operator.
The current implementation, which will be described in detail elsewhere~\cite{guidon:inprep}, goes beyond the minimum image convention, and instead
employs a truncated Coulomb operator which is defined as
\begin{equation}
g(r_{12})=\left\{\begin{array}{rl} \frac{1}{r_{12}}, & r_{12}\leq R_c \\ 0, & r_{12}>R_c.\end{array}\right.
\end{equation}
This operator was suggested by Spencer and Alavi~\cite{spencer:prb:08} to obtain rapid 
convergence for the Hartree-Fock energy with respect to the $k$-point sampling
of the exchange energy in periodic systems.
Note that the use of the truncated Coulomb operator implies that the exchange energy is unconditionally convergent for all $k$ points.
Furthermore, since exchange in insulators is effective on shorter range compared to the electrostatic interaction,
results converge exponentially to the Hartree-Fock limit as $R_c$ is increased.
In line with the results presented in Ref.~\onlinecite{spencer:prb:08}, we find that for a cubic cell with edge $L$ and $R_c = L/2$,
converged results of the exchange energy can be obtained using the $\Gamma$ point only.
Of course, this requires that the computational cell is sufficiently large 
so that the $\Gamma$-point approximation is acceptable,
which in turn requires that the extent of the maximally localized Wannier functions is smaller than $L/2$.
Consequently, the exchange energy computed in CP2K is defined as
\begin{equation}
   -\frac12\sum_{i,j}\int\int \psi_i(r)\psi_j(r)g(|r-r'|)\psi_i(r')\psi_j(r')\, d^3r d^3r',
\end{equation}
where ${\psi_i}$ are the wave functions at the $\Gamma$ point.
In the Gaussian basis set employed, the exchange energy per cell is thus obtained from
\begin{equation}
   -\frac12\sum_{\mu \nu \gamma \delta}\sum_{\mathbf{a}\mathbf{b}\mathbf{c}} P^{\mu\gamma}P^{\nu \delta}(\mu \nu^{\mathbf{a}}|\gamma^{\mathbf{b}}\delta^{\mathbf{b}+\mathbf{c}})_g,
\end{equation}
where $\mu, \nu, \gamma, \delta$ are the indices of the basis functions in the central cell, and $\mathbf{a, b, c}$ run over all image cells.
Due to the rapid decay of the basis functions, sums over image cells $\mathbf{a}$ and $\mathbf{c}$ converge quickly.
The sum over $\mathbf{b}$ converges quickly and unconditionally for our choice of $g(r_{12})$.
Further technical details, including how to compute efficiently and accurately the required four center integrals $(\mu \nu^{\mathbf{a}}|\gamma^{\mathbf{b}}\delta^{\mathbf{b+c}})_g$ will
be presented elsewhere~\cite{guidon:inprep}.

\section{Results\label{sec:res}}

\subsection{HF energy of LiH at experimental volume\label{sec:hf-exp-vol} using Pad{\'e} approximants}

\begin{figure}[tb]
\includegraphics[width=.99\columnwidth,clip=true]{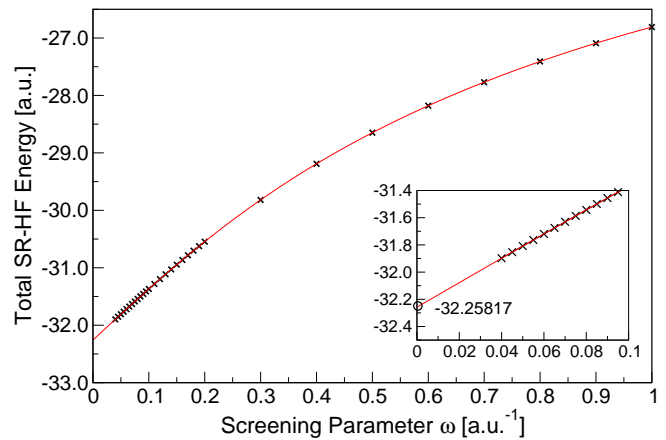}
\caption{\label{fig:ene-om} (Color online) Convergence of the SR-HF energy 
[a.u.] of a LiH unit cell containing four LiH ion pairs at experimental lattice constant (4.084 \AA) with decreasing
screening parameter $\omega$ involved in the short-range Coulomb interaction. Gaussian results for each $\omega$ are
represented by crosses. The line shows the [7/7] Pad{\'e} fit to the numerical data (see text for details). 
The inset gives SR-HF energies
for $\omega \in [0\,; 0.1]$ a.u.$^{-1}$ as well as the extrapolated value for $\omega = 0$ in Hartree atomic units.}
\end{figure}

Fig.~\ref{fig:ene-om} depicts the obtained series of 31 data points of screened HF energies for
various values of $\omega \in [0.040\,;\,1.0]$ (see Sec.~\ref{sec:comp}) calculated at the experimental
lattice parameter. The series of calculated energies
clearly converges to a certain limit with decreasing $\omega$. At this point we remind the reader that
for the limit $\omega \rightarrow 0$ the SR Coulomb kernel given in Eq.~\ref{eq:scr-coul} approaches the
full-range $1/r$ operator. As shown in Fig.~\ref{fig:ene-om} and outlined in Sec.~\ref{sec:comp}, the density
of data points increases significantly toward the $\omega \rightarrow 0$ limit.
Tab.~\ref{tab:fit} presents results for several least-squares fits obtained using rational polynomials up
 to order seven. In addition, correlation coefficient $r$, root-mean-square-deviation (RMSD) as well as
 relative RMSD, which is normalized to the range of observed data, {\it i.e.}~calculated energies, are shown.
Since $r$ is very close to $1$, we decided to present in Tab.~\ref{tab:fit} $(1-r)$, where a value of zero  means perfect 
agreement between calculated data points and fit.

Apparently, the RMSD as well as relative RMSD values decrease with increasing order of the rational polynomial applied
to the fit.
Rational polynomials of order eight or beyond (not shown in
Tab.~\ref{tab:fit}) lead to unstable fits and the goodness of the fit deteriorates.
According to Tab.~\ref{tab:fit} the optimal order of the Pad{\'e} approximant is $[7/7]$, which was used
for all extrapolations employed in this work.

\begin{table}[tb]
\renewcommand{\arraystretch}{1.1}
\setlength{\tabcolsep}{7pt}
\caption{\label{tab:fit} Results for the Pad{\'e} fits to the 31 SR-HF energies [a.u.] of LiH
at experimental lattice constant (4.084 \AA) for a cell containing four LiH ion pairs.
The first column shows the order of the Pad{\'e} polynomials representing them by the order of the polynomial of
the numerator and denominator, respectively (in squared brackets). The extrapolated total HF energy for the cell is
given in Hartree atomic units.}
\begin{threeparttable}
\begin{tabular}{lcccl}
\hline\hline
Fit & (1-r)\tnote{(a)} & RMSD\tnote{(b)} & RMSD\%\tnote{(c)} & E(HF) [a.u.]\\
\hline
$[1/1]$   & 6.7e{-05} &      0.0185955  &      0.3654035 &     -32.3526129 \\
$[2/2]$   & 5.7e{-08} &      0.0005666  &      0.0111347 &     -32.2472782 \\
$[3/3]$   & 5.5e{-07} &      0.0018244  &      0.0358500 &     -32.2521383 \\
$[4/4]$   & 2.0e{-10} &      0.0000364  &      0.0007156 &     -32.2585728 \\
$[5/5]$   & 7.9e{-10} &      0.0000759  &      0.0014909 &     -32.2588628 \\
$[6/6]$   & 1.5e{-09} &      0.0001109  &      0.0021803 &     -32.2576031 \\
$[7/7]$   & 6.2e{-15} &      0.0000002  &      0.0000046 &     -32.2581712 \\
\hline\hline
\end{tabular}
\begin{tablenotes}
\item[(a)] r: correlation coefficient (see text for details). 
\item[(b)] RMSD: root mean square deviation.
\item[(c)] RMSD\%: normalized root mean square deviation.
\end{tablenotes}
\end{threeparttable}
\end{table}

\begin{table}[tb]
\renewcommand{\arraystretch}{1.1}
\setlength{\tabcolsep}{2pt}
\caption{\label{tab:fit-pred}Validation of the [7/7] Pad{\'e} fit. The first column gives
the number of data points ({\it i.e.}~SR-HF energies) included for a [7/7] Pad{\'e} fit in order
to predict the SR-HF energy corresponding to the $\omega$ value given in the second column.
Deviations between calculated and fitted SR-HF energies are given in the fifth and sixth column, 
respectively.}
\begin{ruledtabular}
\begin{tabular}{cccccc}
\#dp  & {$\omega$}  &  E$^{\rm SR-HF}$ [a.u.] &  E$^{\rm fit}$ [a.u.]      &   Error        & \%-Error \\
\hline
 24   &       0.070 &  -31.630779 &   -31.630965 &     -0.0001817 &   \phantom{-}0.000574 \\
 25   &       0.065 &  -31.675185 &   -31.675183 &      \phantom{-}0.0000010 &  -0.000003 \\                                                     
 26   &       0.060 &  -31.719533 &   -31.719533 &      \phantom{-}0.0000003 &  -0.000001 \\                                                     
 27   &       0.055 &  -31.763998 &   -31.763998 &      \phantom{-}0.0000003 &  -0.000001 \\
 28   &       0.050 &  -31.808571 &   -31.808571 &      \phantom{-}0.0000003 &  -0.000001 \\
 29   &       0.045 &  -31.853244 &   -31.853244 &      \phantom{-}0.0000003 &  -0.000001 \\
 30   &       0.040 &  -31.898007 &   -31.898007 &      \phantom{-}0.0000004 &  -0.000001 \\
\end{tabular}
\end{ruledtabular}
\end{table}

As a next step we had to validate the $[7/7]$ polynomial, since it is well known, that algorithms
for interpolation are straightforward, whereas for extrapolation care must be taken. 
A plausible strategy is simply the prediction of energies for a certain value of $\omega$
not included in the fit. 
Table~\ref{tab:fit-pred}
shows a series of predictions for screened HF energies for a series of $\omega$'s starting from
$\omega = 0.070$ a.u.$^{-1}$.
The corresponding screened HF energy 
has been estimated based on a $[7/7]$ fit using 24 data points, where $\omega \in [0.075\,;\, 1.0]$.
As can be seen from Tab.~\ref{tab:fit-pred}, 
it is remarkable that the resulting error is only one order of magnitude larger than the applied SCF 
convergence criterion (see Sec.~\ref{sec:comp}). The error for the predicted energies is practically converged after 
inclusion of only one further data point and amounts to $3\times 10^{-7}$ a.u..
Hence, the error incurred by the fit to the Pad{\'e} approximant is much lower than the convergence threshold in the SCF
procedure. 
By virtue of the aforementioned validations it is safe to give the total HF energy for a unit cell containing four
LiH ion pairs at experimental lattice constant (4.084 \AA) 
with a precision of five decimals in Hartree atomic units, which amounts to $-32.25817$ a.u.. The total HF energy
per formula unit at experimental lattice constant is given in Tab.~\ref{tab:res-bm} and compared with the HF energy obtained
using CP2K. Both values agree excellently to significant precision.

\begin{figure}[tb]
\includegraphics[width=.99\columnwidth,clip=true]{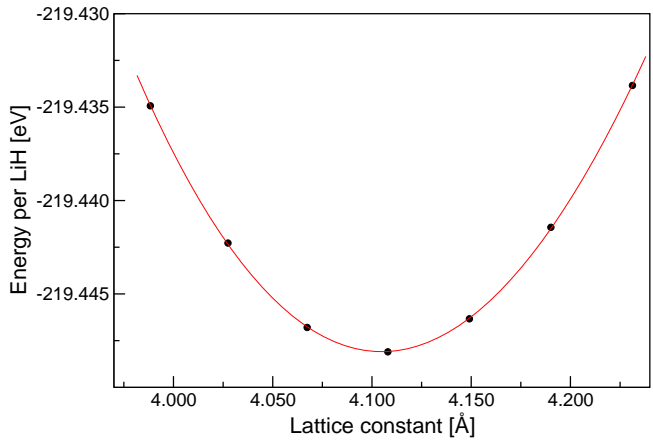}
\caption{\label{fig:hf-scan} (Color online) Murnaghan equation of state (red/gray line) for LiH obtained using the HF
approximation. Each of the seven points corresponds to the extrapolated least-squares fit of 
31 screened HF energies to a Pad{\'e} approximant of order [7/7] (see text for details).}
\end{figure}

\begin{table}[tb]
\renewcommand{\arraystretch}{1.1}
\setlength{\tabcolsep}{5pt}
\caption{\label{tab:res-bm} 
Summary of total HF energies per formula unit, HF cohesive energies, equilibrium lattice constants and bulk moduli of LiH
obtained using GAUSSIAN and CP2K. For comparison purpose, results found in the literature are included.
}
\begin{threeparttable}
\begin{tabular}{lclcc}
\hline\hline
                        & E(HF) [E$_h$] & $\varepsilon_{\rm HF}^{\rm coh}$ [mE$_h$]    &   $a_0$ [\AA]           & B [GPa]          \\
\hline
GAUSSIAN                &   -8.064543$^{\rm a}$    &                              &   4.105                 & 32.34            \\
CP2K                    &   -8.064545$^{\rm a}$    &   -131.949$^{\rm a}$         &                         &                  \\
CRYSTAL$^{\rm b}$       &                          &   -129.14                    &   4.121                 & 28.3\phantom{0}  \\
CRYSTAL$^{\rm c}$       &                          &   -130.16                    &                         &                  \\
VASP$^{\rm d}$          &                          &   -131.7$^{\rm a}$           &                         &                  \\
Gillan {\it et al.}$^{\rm e}$     &                          &   -131.95$^{\rm a}$          &                         &                  \\
Gillan {\it et al.}$^{\rm e}$     &                          &   -131.99                    &   4.108                 & 32.05            \\
\hline\hline
\end{tabular}
\begin{tablenotes}[para]
\item[a] calculated at experimental lattice constant (4.084 \AA).
\item[b] Ref.~\onlinecite{casassa:tca:07}.
\item[c] Ref.~\onlinecite{dovesi:prb:84}.
\item[d] Ref.~\onlinecite{marsman:jcp:09}.
\item[e] Ref.~\onlinecite{gillan:jcc:08}.
\end{tablenotes}
\end{threeparttable}
\end{table}

\subsection{HF lattice constant and bulk modulus with GAUSSIAN}

Fig.~\ref{fig:hf-scan} shows the seven-points-fit of the obtained Pad{\'e} extrapolated HF energies to the Murnaghan equation of state as 
outlined in Sec.~\ref{sec:srHF}. The RMSD value of this fit amounts to $1.4 \times 10^{-4}$.
The resulting HF equilibrium lattice constant of LiH equals 4.105~\AA\ and is in excellent agreement with the result obtained by 
Gillan {\it et al.} (see Tab.~\ref{tab:res-bm}). The corresponding bulk modulus of LiH amounts to 32.34~GPa, which is again
in very good agreement (0.9\% deviation) with the results obtained by aforementioned workers. Note that bulk moduli are quite
sensitive to the equilibrium volume at which they are evaluated and overall good indicators for the quality of the underlying
energies at the various volumes.

\subsection{Total and cohesive energy at experimental volume with CP2K}
Total energies have been computed for systems of increasing system size by explicitly repeating the cubic unit cell periodically in three dimensions.
The largest cell employed is a $5 \times 5 \times 5$ repetition of the basic cubic cell, and contains exactly 1000 atoms.
For this system, $37500$ Gaussian basis functions are used for the expansion of the molecular orbitals, which makes this a computationally demanding simulation.
With increasing system size, we also increase the range of the truncated Coulomb operator, in steps of 2~\AA{} up to a maximum of 10~\AA{} (see Tab.~\ref{tab:cp2k-etot}).
The $\Gamma$-point approximation therefore converges quickly (exponentially) to the HF limit of this system.
We thus obtain from a direct calculation, without extrapolation, an accurate estimate of the total energy per unit cell of approximately -32.258179 a.u..
The finite size error on this result is estimated to be smaller than 50 $\mu$E$_h$.
Furthermore, this number is in excellent agreement with the Pad\'e-extrapolated SR-HF results (-32.258171 a.u., Tab.~\ref{tab:fit}),
and thus provides numerical evidence for the quality of both approaches.
Calculating the HF energy of the H atom and the Li atom with the current basis set,
in periodic boundary conditions and retaining the basis functions of all other atoms in the unit cell,
we can obtain a consistent estimate of the cohesive energy.
In our approach, due to the fact that unrestricted calculations are needed for the atoms, 
these calculations are even more demanding than the bulk,
and have only been performed up to a $4\times4\times4$ repetition of the basis unit cell.
Our best estimate for the cohesive energy, obtained from just three calculations (bulk LiH, and the atoms Li, H) without extrapolation, is -131.949 mE$_h$.
Also here, the finite size error is estimated to be smaller than 50 $\mu$E$_h$.
This number is in excellent agreement with the best estimate obtained by Gillan et al.~\cite{gillan:jcc:08} -131.95 mE$_h$.

\begin{table}[tb]
\renewcommand{\arraystretch}{1.1}
\setlength{\tabcolsep}{3pt}
\caption{\label{tab:cp2k-etot} Results obtained with CP2K and the truncated Coulomb operator for unit cells that are a multiple of the cubic unit cell (4.084 \AA).
The columns show the size of the unit cell, the range of the truncated Coulomb operator ($R_c$),
the Hartree-Fock energy per four LiH ion pairs, the H atom energy, the Li atom energy, and the cohesive energy ($\varepsilon_{HF}^{coh}$), respectively.
}
\begin{threeparttable}
\begin{tabular}{lccccl}
\hline\hline
       & $R_c$[\AA{}] & E(HF)[a.u.] & H[a.u.]\tnote{(a)}  &        Li[a.u.]\tnote{(b)}  &  $\varepsilon_{\rm HF}^{\rm coh}$ [a.u.] \\
\hline
$2\!\times\!2\!\times\!2$  &   4.0 &  -32.244609 & -0.499957 & -7.428493  &   -0.132702  \\
$3\!\times\!3\!\times\!3$  &   6.0 &  -32.256844 & -0.499974 & -7.432137  &   -0.132100  \\
$4\!\times\!4\!\times\!4$  &   8.0 &  -32.258022 & -0.499974 & -7.432582  &   -0.131949  \\
$5\!\times\!5\!\times\!5$  &  10.0 &  -32.258179 &  N/A      &     N/A    &       N/A   \\
\hline\hline
\end{tabular}
\begin{tablenotes}
\item[(a)] basis set limit -0.500000
\item[(b)] basis set limit -7.432727
\end{tablenotes}
\end{threeparttable}
\end{table}

\section{Conclusions\label{sec:conc}}
The Hartree-Fock energy of solid LiH has been calculated using large Gaussian basis sets. Two different
approaches, extrapolation of a Pad{\'e} fit to a series of SR-HFX calculations and direct calculation using a
truncated Coulomb operator, have been found to yield total energies that agree to better than 0.1 mE$_h$.
Calculations of the cohesive energy, the equilibrium lattice constant and the bulk modulus agree with the best
estimates available in literature. These results show that robust and accurate calculations with nearly converged
Gaussian basis sets have now become possible in the condensed phase at least for large band gap systems. These results
will contribute to the growing usefulness of hybrid density functionals for condensed phase applications and
opens, for these systems, the way to accurate calculations based on post-Hartree-Fock methods.

\section*{Acknowledgments}
J.V. acknowledges a fruitful collaboration with John Levesque (Cray Inc.).
His extensive testing on the Cray XT5 (Jaguar) at the Oak Ridge National Laboratory (ORNL) enabled the massively parallel implementation of the Hartree-Fock code in CP2K.
Additional computer resources have been provided by the Swiss National Supercomputing Center (CSCS), and as part of the PRACE project by the Finnish IT Center for Science (CSC).
The PRACE project receives funding from the EU's Seventh Framework Programme (FP7/2007-2013) under grant agreement no RI-211528.
This work was supported by DOE DE-FG02-04ER15523, and the Welch Foundation C-0036.

\end{document}